\documentclass[12pt, twoside]{article}
\usepackage[a4paper,left=2cm,right=2cm,top=2.5cm]{geometry}
\usepackage{setspace}
\usepackage{natbib}
\usepackage{hyperref}
\usepackage[utf8]{inputenc}
\usepackage{url}
\usepackage{bm}
\usepackage{amsmath}
\usepackage{todonotes}
\usepackage{float}
\usepackage{graphicx}
\usepackage{subcaption}
\usepackage{booktabs}
\usepackage{fancyhdr}
\usepackage{algorithm}
\usepackage[noend]{algpseudocode}
\def\BState{\State\hskip-\ALG@thistlm}
\newcommand\Algphase[1]{%
	\vspace*{-.7\baselineskip}\Statex\hspace*{\dimexpr-\algorithmicindent-2pt\relax}%\rule{\textwidth}{0.4pt}%
	\Statex\hspace*{-\algorithmicindent}\textbf{#1}%
	\vspace*{-.7\baselineskip}\Statex\hspace*{\dimexpr-\algorithmicindent-2pt\relax}%\rule{\textwidth}{0.4pt}%
}
\fancypagestyle{plain}{%
	\fancyhf{}
	\fancyfoot[C]{\iffloatpage{}{\thepage}}
	}
\graphicspath{{Figures/}}
\title{RaJIVE: Robust Angle Based JIVE for Integrating Noisy Multi-Source Data}
\author{Erica Ponzi$^1$, Magne Thoresen$^1$, Abhik Ghosh$^2$\\
\small{1. Oslo Center for Biostatistics and Epidemiology, UiO, University of Oslo,  Norway} \\
\small{2. Interdisciplinary Statistical Research Unit, Indian Statistical Institute, Kolkata, India} 
}
\date{}

\begin{document}

	\maketitle
\begin{abstract}
	\noindent
	\textbf{Motivation:} With increasing availability of high dimensional, multi-source data, 
	the identification of joint and data specific patterns of variability has become a subject of interest in many research areas. 
	Several matrix decomposition methods have been formulated for this purpose, 
	for example JIVE (Joint and Individual Variation Explained), and its angle based variation, aJIVE. 
	Although the effect of data contamination on the estimated joint and individual components has not been considered in the literature, 
	gross errors and outliers in the data can cause instability in such methods, 
	and lead to incorrect estimation of joint and individual variance components.  \\
	\textbf{Results:}  
	We focus on the aJIVE factorization method and provide a thorough analysis of the effect outliers on the resulting variation decomposition. 
	After showing that such effect is not negligible when all data-sources are contaminated, 
	we propose a robust extension of aJIVE (RaJIVE) that integrates a robust formulation of the singular value decomposition into the aJIVE approach. 
	The proposed RaJIVE is shown to provide correct decompositions even in the presence of outliers and improves the performance of aJIVE. 
	We use extensive simulation studies with different levels of data contamination to compare the two methods. 
	Finally, we describe an application of RaJIVE to a multi-omics breast cancer dataset from The Cancer Genome Atlas. \\
	\textbf{Availability and implementation:} We provide the \texttt{R} package \texttt{RaJIVE} with a ready-to-use implementation of the methods and documentation of code and examples. \\
	\textbf{Contact:} \href{erica.ponzi@medisin.uio.no}{erica.ponzi@medisin.uio.no}\\
\end{abstract}

\section{Introduction}

The analysis of high dimensional data has recently become a major statistical challenge in many different fields of scientific research, 
where advanced technologies and computing resources enable the collection of extremely large datasets.
In several cases, data are collected from different measurement devices and techniques. 
For example, in genomic studies, multiple layers of omics data are collected on the same organisms or tissues, and measured on different platforms. 
Integration of these high-dimensional data, obtained from multiple sources, is required for their simultaneous analyses
which has received growing interest in recent times \citep{tseng.etal2015, huang.etal2017, rappaport.ron2018}. 

In such multi-source data, it is important to identify the variation patterns 
that are shared among the different data layers, arising from the underlying  joint processes, 
and to separate them from the individual, source specific components of variation. 
Proper identification of both types of variation components can contribute to a more accurate interpretation of the underlying  processes 
and yield better model predictions \citep{mage.etal2019, ponzi.etal2020}.
Several methods have been proposed for this purpose of separating the common and the component-wise variations from multiple data sources 
 \citep{lofsted.etal2012, lock.etal2013, schouteden.etal2013, FENG2018241, tang2018integrated, fan.etal2019}; 
most of them are based on a common framework of matrix decomposition. 
Among these, JIVE (Joint and Individual Variation Explained) \citep{lock.etal2013} has become widely popular, 
especially in medical applications \citep{hellton.thoresen2016, kuligowski.etal2015, kaplan.lock2017}. 
JIVE minimizes the squared residual components of the decomposition of the data matrix, 
using an iterative algorithm that alternatively estimates the joint and individual components using singular value decomposition (SVD). 
Subsequently, \cite{FENG2018241} proposed an angle based variation of JIVE (aJIVE), 
which computes the matrix decomposition by using perturbations of the row spaces and 
is computationally more efficient than the original JIVE.

Although aJIVE improves JIVE by reducing its computational burden, it still relies on the SVD for factorization of 
the data matrices into joint, individual and residual components. 
Usual computation of SVD is, however, highly sensitive to the presence of any gross errors or outliers in the data 
\citep{zhang.etal2013,Hawkins.etal2001}.  
As a consequence, both JIVE and aJIVE become extremely unstable in presence of noise in the data 
and often lead to incorrect inference on the resulting joint and individual variation components. 
But, modern large-scale datasets collected via different sources, e.g., multi-omics data, 
are prone to different kinds of noise and outlying observations, and hence, 
an appropriate robust algorithm is of utmost practical necessity to integrate such multi-source datasets 
for stable inferential conclusion.

Surprisingly, the existing literature on this important topic is very limited. 
A robust alternative to JIVE has recently been proposed \citep{sagonas.etal2017} 
with successful applications in image recognition and facial analysis. 
However, since it does not utilize the idea of aJIVE to reduce the computational burden, 
it becomes even more computationally challenging than the original JIVE for large-scale datasets. 
Another alternative procedure for data integration, 
namely the Robust Correlated and Individual Components Analysis \citep{panagakis.etal2016}, 
provides a robust solution against outliers in the data; 
but it is limited to the analysis of two datasets only, and does not generalize for integration of more data-sources. 
Further, both methods have only been applied to image recognition problems  
but no biological application is found in the literature.

This present work focuses on robustly identifying the common and the distinct variation components from multi-source data 
in the presence of outliers in either (or some) of the data sources  ($\geq2$). 
We propose a novel robust extension of the aJIVE algorithm, which we will refer to as ``RaJIVE", 
by integrating a robust formulation of the SVD within the computationally efficient angle based approach of aJIVE. 
%
%In this work, we incorporate the robust SVD formulation into the aJIVE framework, to develop 
%a novel efficient algorithm for robust integration of noisy multi-source datasets; 
We assess the performance of the proposed RaJIVE method via extensive simulation studies, 
where we investigate the correct recovery of individual and joint components in the presence of 
increasing proportions of outliers within appropriate synthetic datasets. 
Additionally, considering different outlier configurations, we also illustrate their effects on the standard aJIVE decomposition. 
To our knowledge, the effects of outliers on such decomposition of joint and individual variation components 
are not yet investigated in the literature; 
our thorough analyses of the consequences of different types of noises would be highly beneficial for a large class of multi-source studies. 
The proposed RaJIVE is shown to provide significantly improved results compared to the usual aJIVE under data contamination.

Finally, we describe a real-life application to multi-source omics data from The Cancer Genome Atlas 
(TCGA Research Network \href{https://www.cancer.gov/tcga}{https://www.cancer.gov/tcga}), 
where we apply RaJIVE to simultaneously analyze methylation, gene expression and miRNA data collected on breast cancer samples.

%\newpage
\section{Methods}
\subsection{Angle Based JIVE (aJIVE)}
\label{SEC:aJIVE}

In a data integration setup with $K$ data-sources and $n$ subjects, each data-block is denoted as $\bm{X}_k$, $k =1, ..., K$,
which is a matrix with $n$ columns for $n$ subjects and  $p_k$ rows corresponding to the variables in the $k$-th data-source. 
The overall dimensionality of the data is then $p = p_1 +...+p_K$.
Our aim is to identify the joint variation components among these $K$ data-matrices 
and separate them from the individual source-specific variation (and residual)  components.  
Mathematically, we consider the decomposition of each data-matrix into the sum of three low-rank terms as follows. 
\begin{align}
\bm{X}_1 &= \bm{J}_1 + \bm{I}_1 + \bm{\epsilon}_1 \notag \\
&\;\;\vdots \label{decomp} \\
\bm{X}_K &= \bm{J}_K + \bm{I}_K + \bm{\epsilon}_K. \notag 
\end{align}
where $\bm{I}_k$ is the individual component for the $k$-th data-block, $\bm{\epsilon}_k$ is its residual (random error) component and 
\begin{equation}
\bm{J} =  \begin{bmatrix}
\bm{J}_1 \\
... \\
%... \\
\bm{J}_K
\end{bmatrix}
\end{equation}
%\bm{J}$
is the joint structure matrix with each $\bm{J}_k$ being the submatrix  associated with $\bm{X}_k$.   
We then need to identify (estimate) these submatrices $\bm{J}_k$ and $\bm{I}_k$, $k=1, \ldots, K$, 
from the observed data-matrices $\bm{X}_k$.

The  aJIVE method \citep{FENG2018241} is a popular algorithm for this data-integration problem, which  is structured in three phases: 
First it computes the low-rank approximation of each data block $\bm{X}_k$, by using its SVD. 
In the second phase, called score space segmentation, it extracts the joint structure between the obtained low-rank approximations, 
by computing the SVD of the stacked row basis matrices, based on the principles of \textit{Principal Angle Analysis}. 
Finally, each data block is projected onto the joint bases to obtain the joint components $\bm{J}_k$;
the individual components $\bm{I}_k$ are then calculated by orthonormal basis subtraction.
Ready-to-use implementation of aJIVE is available in both \texttt{Matlab} \citep{ajiveM} and \texttt{R} \citep{ajiveR}.

\subsection{The Proposed Algorithm: RaJIVE}

In order to correct the non-robust nature of aJIVE, we propose a new novel RaJIVE algorithm
where we incorporate a robust SVD approach of \cite{zhang.etal2013} within the aJIVE framework of data integration. 
There are two existing approaches of robust SVD computation \citep{Hawkins.etal2001,zhang.etal2013};
both use a robust loss for the minimization of the error between the data matrix and the SVD reconstruction
after formulating the problem to an alternating (linear) regression procedure.
Specifically, the robust $\ell_1$-loss is used in \cite{Hawkins.etal2001} and \cite{zhang.etal2013} used 
the Huber loss associated with M-estimation of the regression coefficients \citep{huber.ronchetti2011}. 
This Huber loss is known to yield more efficient estimator of regression coefficients compared to the $\ell_1$-loss \citep{huber.ronchetti2011}
and also the corresponding robust SVD computation requires lower computation time. 
For these reasons, after investigating both robust SVD approaches, 
we propose to use the one from \cite{zhang.etal2013} which additionally includes a suitable provision for allowing missing values in the datasets. 
This particular robust SVD algorithm, to be referred to as the robRSVD, 
is devised based on the idea of subsequently using iterative reweighted least-squares to compute the regression coefficients at each step. 
The weights are computed based on the Huber's function \citep{RobRpackage}  
to achieve robustness of the resulting singular values and the associated singular vectors. 
For the sake of completeness, a brief discussion about the computation of this robRSVD is presented in Appendix \ref{App1}.

%\textbf{Algorithm 1: [pseudo-code for RaJIVE]}:  
\begin{algorithm}[!b]
	\caption{Pseudo-code for RaJIVE}\label{rAjive.alg}
	\hspace*{\algorithmicindent} \textbf{Input}: Data {$\bm{X}_k$, $k = 1, ..., K$}, initial individual ranks $r_1, ..., r_K$ 
	\\
	\hspace*{\algorithmicindent} \textbf{Output}: Joint $\bm{J}_k$, Individual $\bm{I}_k$ and Noise $\bm{E}_k$ Components
	\begin{algorithmic}[1]
		%	\Procedure{RaJIVE} {} 
		\Algphase{Phase 1: Initial Signal Space Extraction}
		%\State \textbf{}
		\For{\texttt{$k = 1, ..., K$}} %\hfill \COMMENT{In Parallel}
		\State \texttt{Compute the robust SVD of block $\bm{X}_k$
			\State \texttt{Threshold based on initial rank specifications}: $\widetilde{\bm{U}}_k\widetilde{\bm{\Sigma}}_k\widetilde{\bm{V}}^T_k$}
		
		\EndFor 
		\Algphase{Phase 2: Score Space Segmentation}
		\State $\bm{M} \gets  \begin{bmatrix}
		\widetilde{\bm{V}}_1 \\
		... \\
		... \\
		\widetilde{\bm{V}}_K
		\end{bmatrix}$  \hfill Stacked singular vectors matrix
		\State \texttt{Compute the robust SVD of $\bm{M}$ and identify principal angles}
		\Algphase{Phase 3: Final decomposition}
		\For{\texttt{$k = 1, ..., K$}} 
		\State $P_J \gets \hat{V_J}\hat{V_J}^T$ \hfill Projection matrix onto joint space
		\State $J_k \gets  X_k P_J$ \hfill Projection of $X_k$ onto joint basis
		\State $I_k \gets  X_k P_J^\perp$ \hfill Orthonormal basis subtraction
		\State $I_k \gets \texttt{Threshold robust SVD of $I_k$}$
		\State $E_k \gets \texttt{Remaining components of each SVD}$

		\EndFor
		%		\EndProcedure
	\end{algorithmic}
	
\end{algorithm}

Our proposed RaJIVE uses this robRSVD instead of the usual SVD in each step of iterations, but works in three phases as in aJIVE. 
In the first phase, it uses robRSVD to compute the low-rank approximations of each data-matrix $\bm{X}_k$.
Secondly, we stack the row basis matrices obtained from the robRSVD of each block, 
and extract the (robust estimate of) joint-structure by another application of robRSVD to the stacked matrix, 
following the same principle as in aJIVE. 
Once the joint structure is obtained, each data block is projected onto the estimated robust joint-bases to compute the joint components $\bm{J}_k$.
Finally, the robust estimates of the individual $\bm{I}_k$ are computed through orthonormal basis subtraction.
Additionally, we have appropriately modified the implementation of RaJIVE (from that of aJIVE) to allow for parallel computation, 
which significantly reduces its computational time.
The pseudo-code for the final RaJIVE algorithm is presented in Algorithm 1. 
We also provide an \texttt{R} package, \texttt{RaJIVE}, containing the ready-to-use implementation of our proposed algorithm.

%\bigskip
%\noindent

%\bigskip
%\bigskip
%\clearpage

\section{Simulation studies}

\subsection{Simulation Scheme}
\label{sec:sim}

\noindent\textbf{Settings:} $K=3$  data sets were simulated, with $n$ columns each 
and different number of rows as $p_1, p_2, p_3$, respectively. 
The true joint rank was taken as $r$ and the true individual ranks as $r_1, r_2, r_3$, respectively. 
Outliers were generated by adding random errors from $\mathcal{N}(\mu_c, \sigma_c^2)$ to a given proportion of variables and observations,
as per the following six configurations: 
(O1) outliers in $p_k/5$ variables across all three data-sources, 
(O2) outliers in $p_k/5$ variables in the data-source with highest initial rank, 
(O3) outliers in $p_k/5$ variables in the data-source with lowest initial rank, 
(O4) outliers in all $p_k$ variables across the three data sources, 
(O5) outliers in all $p_k$ variables in the data source with highest initial rank and 
(O6) outliers in all $p_k$ variables in the data source with lowest initial rank,
where $p_k$ is the number of variables in the respective $k$-th data-matrix. 
Additionally, different proportions of observations were contaminated in each cases.
The simulation study was repeated 100 times to see the variability in the results.

\noindent\textbf{Performance measures:} 
For each set of simulated data-matrices, the usual aJIVE and the proposed RaJIVE was performed with and without the outliers 
and results were compared in terms of the joint rank, the individual ranks and the proportions of variance explained. 
Additionally, the subspace recovery error was also considered, 
which measures the distance between the true (simulated) and estimated joint subspaces as

\begin{equation}
SRE = \frac{1}{r} ||\bm{\hat{U}}^T\bm{\hat{U}} -\bm{U}^T\bm{U} ||.
\end{equation}
Here, $\bm{\hat{U}} = [\bm{\hat{u_1}}, ..., \bm{\hat{u_r}}]$ and $\bm{U} = [\bm{u_1}, ..., \bm{u_r}]$ are
respectively the estimated and the true (simulated) joint eigenbasis of dimension $r$. 
Finally, binary responses were simulated depending on the true joint scores, 
and a logistic model were fitted to classify this binary response based on the estimated joint scores;
the resulting prediction AUCs were compared as another summary performance measure.

\subsection{The effect of outliers on aJIVE}
\label{sec:aJIVEsim}

We first conducted simulation studies to investigate the effect of outliers on the aJIVE decomposition on synthetic datasets, 
which provided the motivation for developing a robust version of this data-integration algorithm.
To this end, we have considered $n = 100$,  $(p_1, p_2, p_3) = (200, 180, 145)$, $r=2$,  and $(r_1, r_2, r_3)=(20,12,12)$
and outliers were from $\mathcal{N}(10, 4)$. 
For brevity, we report the  results obtained by adding outliers to $10\%$ of the total samples;
the same patten were observed for other outlier proportions as well. 

\begin{figure}
	\centering
	\includegraphics[width=.99\linewidth]{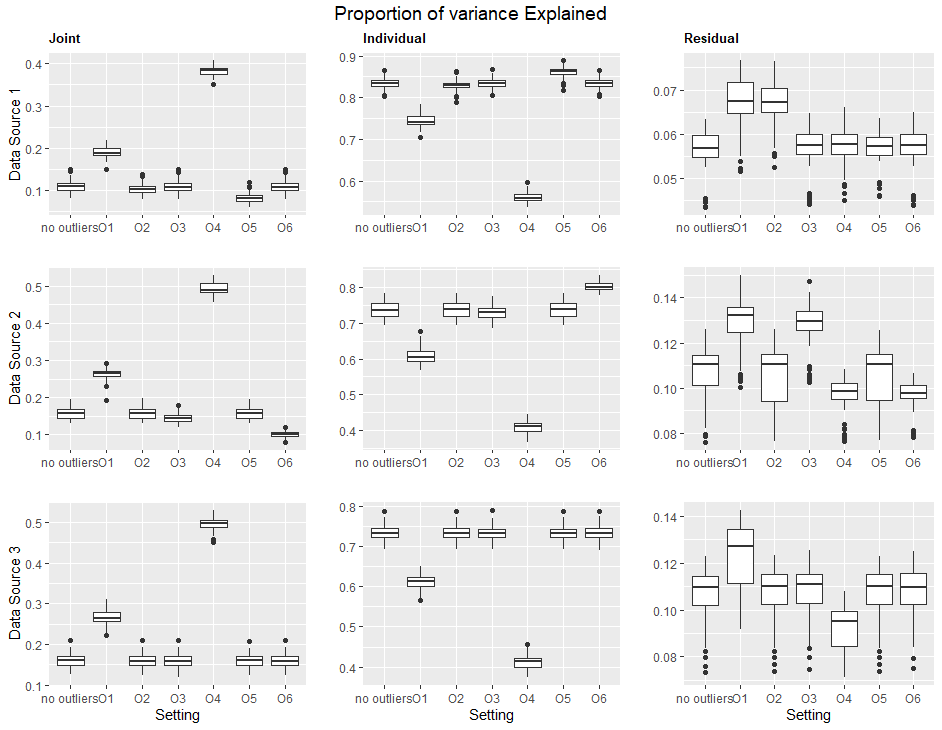}
	\caption{ Proportions of variance explained estimated by aJIVE in a standard setting (no outliers)
		vs in the six simulated outlier configurations O1--O6.}
	\label{fig:outliersV}
\end{figure}

\begin{figure}
	\centering
	\begin{subfigure}{.5\textwidth}
		\centering
		\includegraphics[width=.95\linewidth]{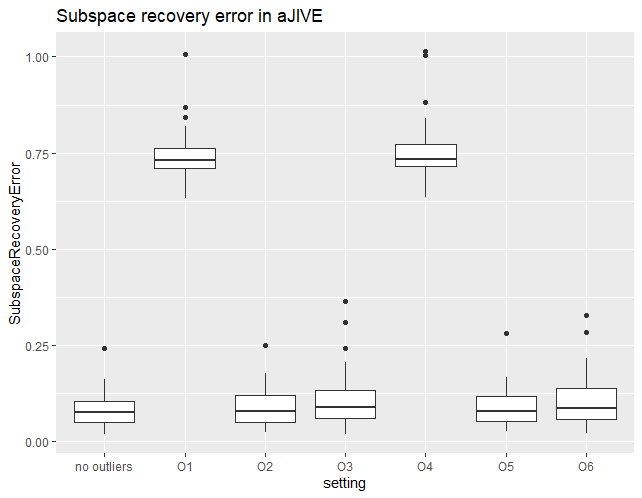}
		\caption{ }
		\label{fig:outliers_sub1}
	\end{subfigure}%
	\begin{subfigure}{.5\textwidth}
		\centering
		\includegraphics[width=.95\linewidth]{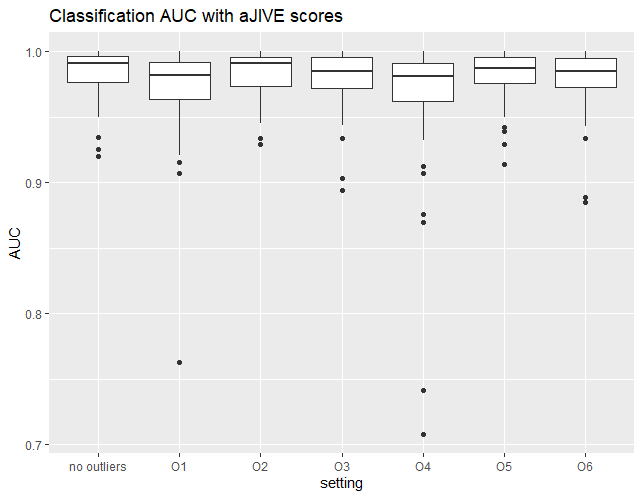}
		\caption{ }
		\label{fig:outliers_sub2}
	\end{subfigure}
	
	\label{fig:outliers}
	\caption{Subspace recovery error and classification AUC from aJIVE scores in a standard setting (no outliers)
		vs in the six simulated outlier configurations O1--O6.}
\end{figure}

Results show that only when outliers are added to all the three data sources (case O1 and case O4), the estimated aJIVE parameters are affected. This was observed both when $p_k/5$ (case O1) and $p_k$ (case O4) variables were contaminated by outliers.
In particular, the joint rank is higher when outliers are present (Table \ref{tab:outliersR}), suggesting that one or more joint components might actually be residual noise. The proportions of variance explained are also affected by outliers, specifically showing higher joint and lower individual proportions (Figure \ref{fig:outliersV}).
The subspace recovery error is substantially higher when outliers are present (Figure \ref{fig:outliers_sub1}), while the AUC does not seem to be affected (Figure \ref{fig:outliers_sub2}).

\begin{table}[h!]
	\centering
	\begin{tabular}{c| c c c c}
		Configuration & Joint & Ind Source 1 & Ind Source 2 & Ind Source 3 \\ \hline
		No outliers & 2 & 18 & 10 & 10 \\
		O1 & 3 & 17 & 9 & 9 \\
		O2 & 2 & 18 & 10 & 10 \\
		O3 & 2 & 18 & 10 & 10 \\ 
		O4 & 3 & 17 & 9 & 9 \\
		O5 & 2 & 18 & 10 & 10 \\
		O6 & 2 & 18 & 10 & 10 \\
	\end{tabular}
	\caption{Median joint and individual ranks estimated by aJIVE in simulation studies without outliers and with  six outlier configurations.}
	\label{tab:outliersR}
	
\end{table}

\subsection{Performance Evaluation of the RaJIVE}

We evaluate the finite-sample performance of the proposed RaJIVE, in comparison with the usual aJIVE, 
via appropriate simulation studies similar to the one discussed in Section \ref{sec:sim}.
Based on the results of Section \ref{sec:aJIVEsim}, here, we consider outliers in all three datasets of size $n=100$ each,
by adding random normal errors to $5\%$ and $10\%$ of the variables in each dataset, and $10\%$ observations for each of these variables. 
In particular, we consider two simulation scenarios -- one (Set-A) similar to those in Section \ref{sec:aJIVEsim} 
and another one (Set B) resembling our real data situation.

\noindent\textbf{Set-A:} We take  $(p_1, p_2, p_3) = (200, 180, 150)$, $r=3$,  and $(r_1, r_2, r_3)=(20,12,7)$.
Outliers were generated from $\mathcal{N}(15, 1)$.

\noindent\textbf{Set-B:} We take  $(p_1, p_2, p_3) = (100, 180, 150)$, $r=3$,  and $(r_1, r_2, r_3)=(10,12,7)$.
Outliers were generated from a normal distribution with mean $\mu_c = 3m+5s$ and standard-deviation $\sigma_c = 3s$, 
where $m$ and $s$ are mean and standard-deviation of the affected variable, respectively.

\begin{figure}
	\centering
	\textbf{\large{Estimated proportions of variance explained}} \\
	\vspace{0.4cm}
	\begin{subfigure}{.8\textwidth}
		\centering
		\includegraphics[width=.95\linewidth]{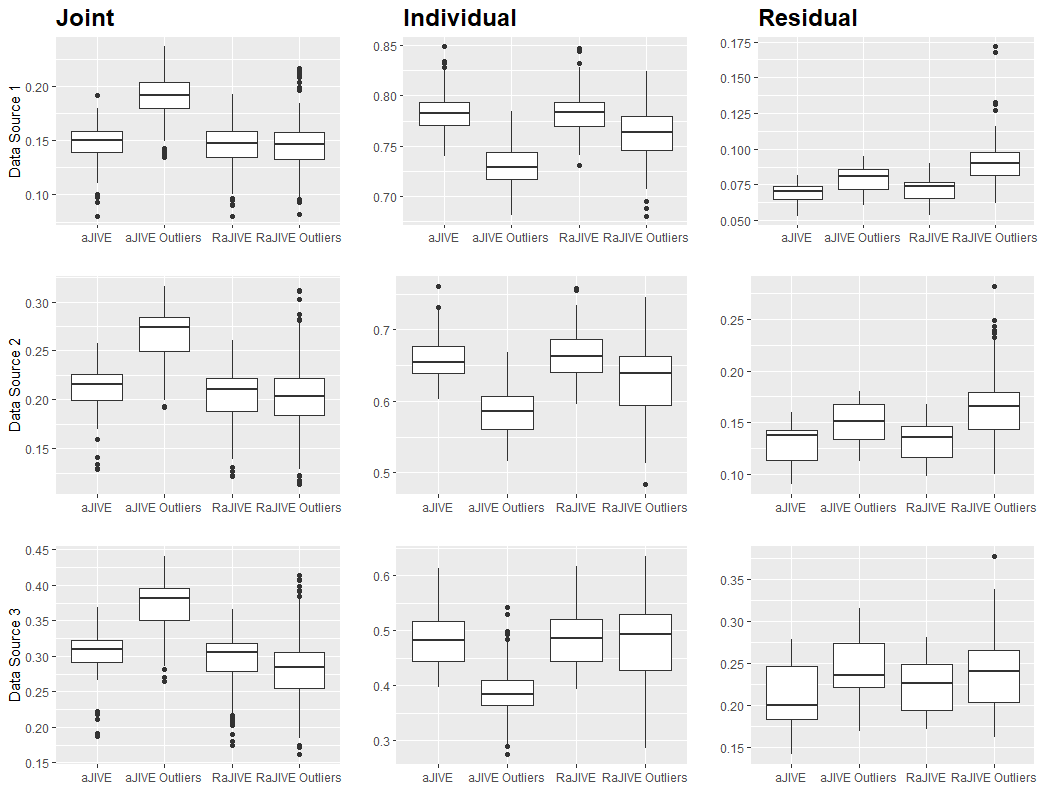}
		\caption{Outliers in $5\%$ of the variables}
		\label{fig:simVar5.10_SetA}
	\end{subfigure}%
%	\begin{subfigure}{.5\textwidth}
%		\centering
%		\includegraphics[width=.99\linewidth]{SimulationVariance5_20.png}
%		\caption{Outliers in $5\%p$ and $20\%n$.}
%		\label{fig:simVar5.20}
%	\end{subfigure}
%	
	\vspace{0.4cm}
	\begin{subfigure}{.8\textwidth}
		\centering
		\includegraphics[width=.95\linewidth]{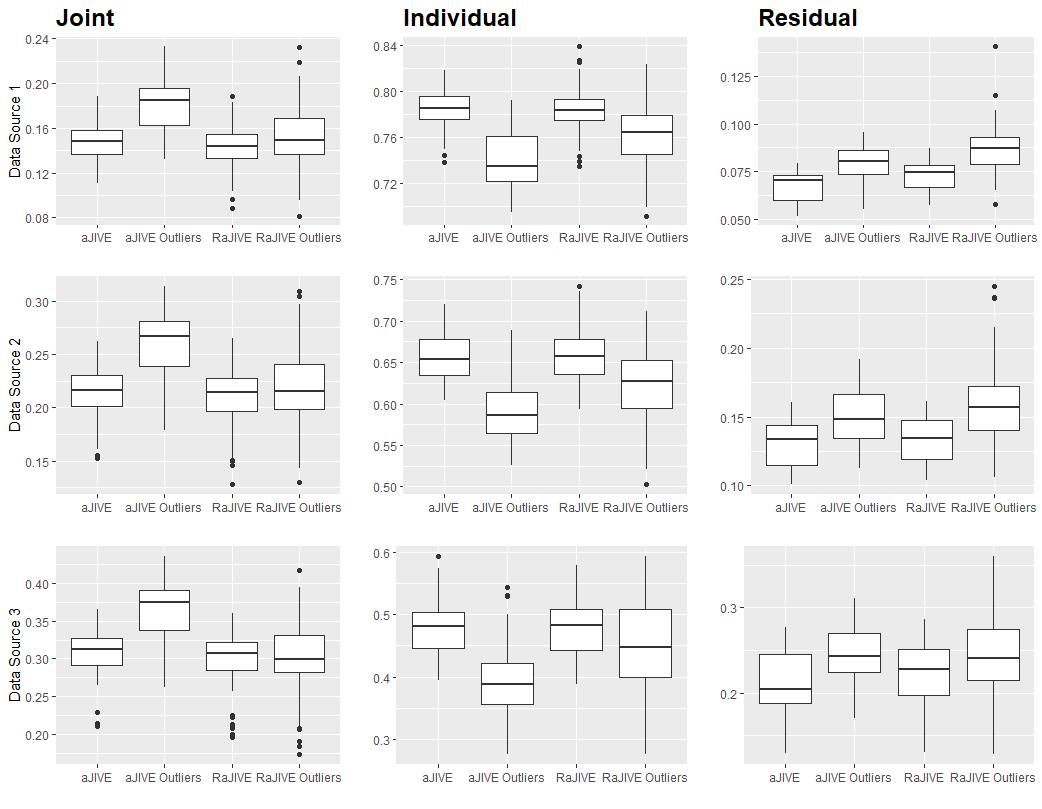}
		\caption{Outliers in $10\%$ of the variables.}
		\label{fig:simVar10_SetA}
	\end{subfigure}
	\caption{Proportions of variance explained, estimated by aJIVE and RaJIVE, 
		in simulation studies (Set A) with outliers in different proportion of variables and $10\%$ samples.}	
	\label{fig:simVar_SetA}
\end{figure}

\begin{figure}
	\centering
	\textbf{\large{Estimated proportions of variance explained}} \\
	\vspace{0.4cm}
	\begin{subfigure}{.8\textwidth}
		\centering
		\includegraphics[width=.95\linewidth]{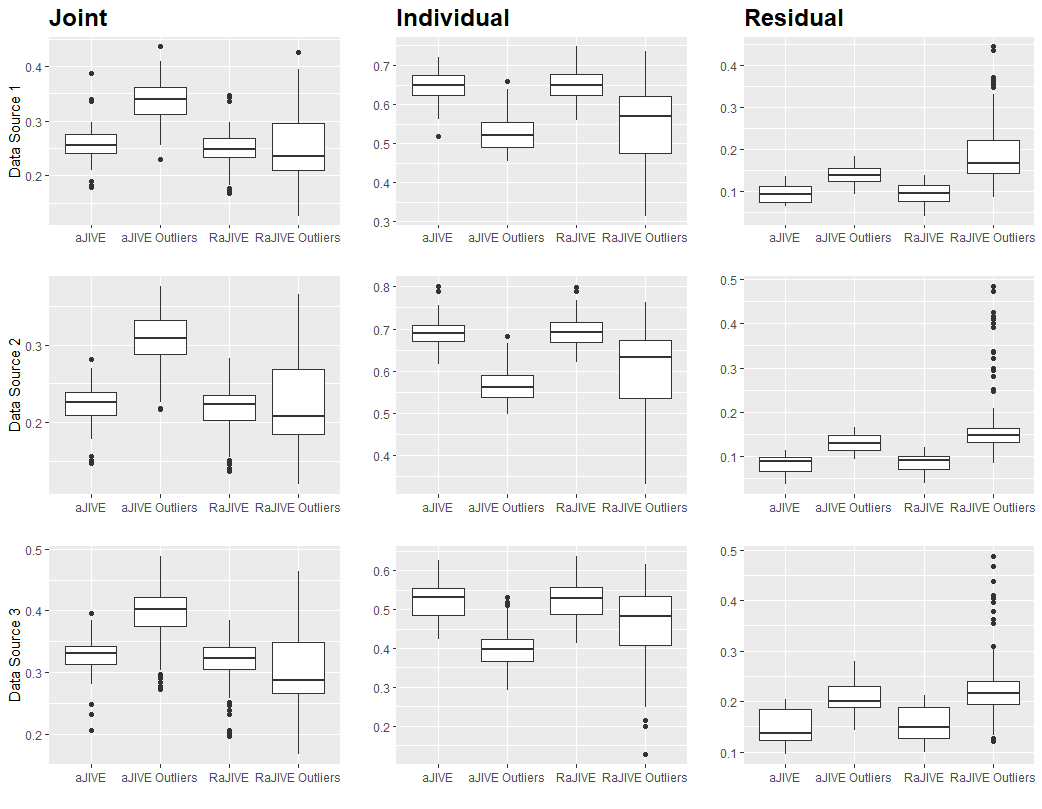}
		\caption{Outliers in $5\%$ of the variables}
		\label{fig:simVar5.10_SetB}
	\end{subfigure}%
	%	\begin{subfigure}{.5\textwidth}
	%		\centering
	%		\includegraphics[width=.99\linewidth]{SimulationVariance5_20.png}
	%		\caption{Outliers in $5\%p$ and $20\%n$.}
	%		\label{fig:simVar5.20}
	%	\end{subfigure}
	
	\vspace{0.4cm}
	\begin{subfigure}{.8\textwidth}
		\centering
		\includegraphics[width=.95\linewidth]{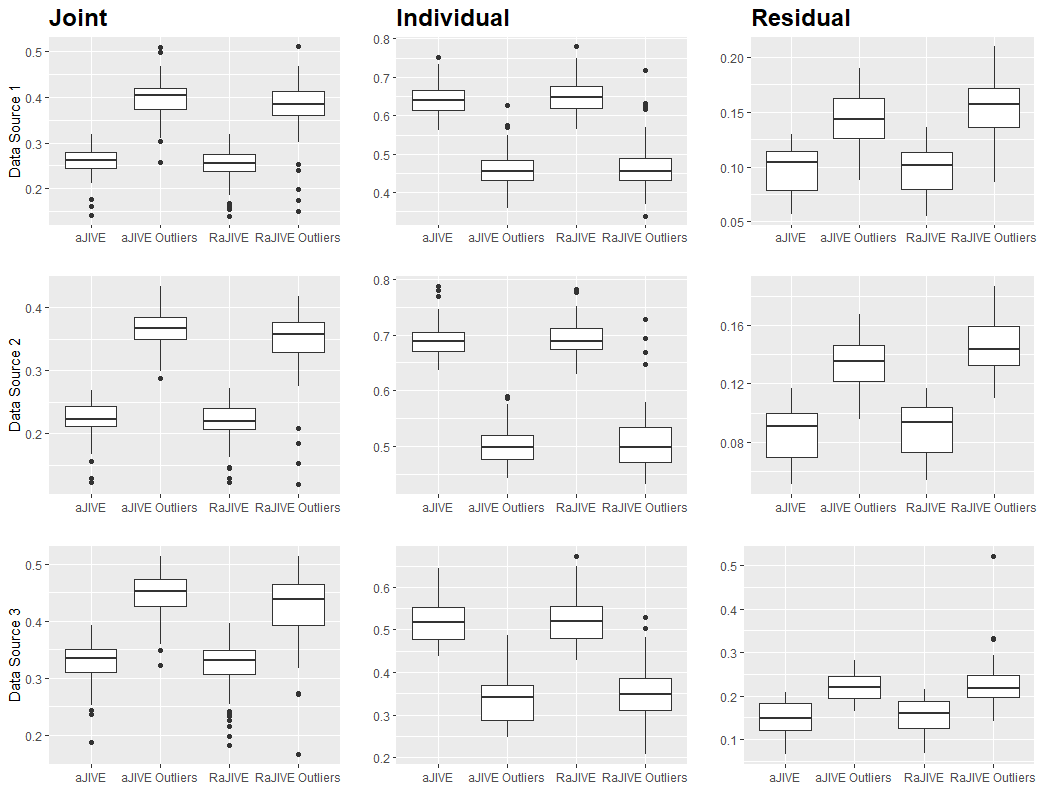}
		\caption{Outliers in $10\%$ of the variables. }
		\label{fig:simVar10_SetB}
	\end{subfigure}%
	%\begin{subfigure}{.5\textwidth}
	%	\centering
	%	\includegraphics[width=.99\linewidth]{SimulationVariance10_20.png}
	%	\caption{Outliers in $10\%p$ and $20\%n$.}
	%	\label{fig:simVar10.20}
	%\end{subfigure}
	\caption{Proportions of variance explained, estimated by aJIVE and RaJIVE, 
		in simulation studies (Set B) with outliers in different proportion of variables and $10\%$ samples.}	
	\label{fig:simVar_SetB}
\end{figure}

\begin{figure}
	\centering
	\textbf{\large{Estimated subspace recovery error}} \\
	\vspace{0.4cm}
	\begin{subfigure}{.5\textwidth}
		\centering
		\includegraphics[width=.99\linewidth]{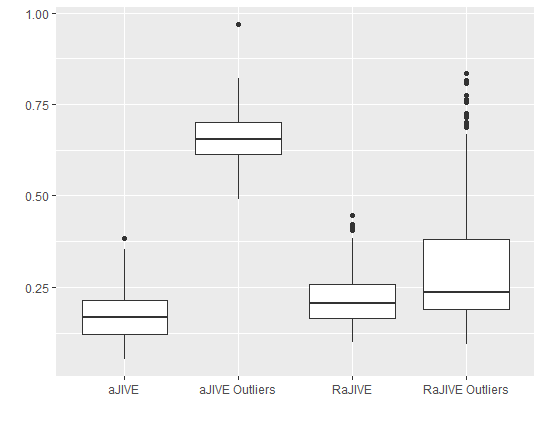}
		\caption{Outliers in $5\%$ variables, Set A.}
		\label{fig:simSRE5.10_SetA}
	\end{subfigure}%
%	\begin{subfigure}{.5\textwidth}
%		\centering
%		\includegraphics[width=.99\linewidth]{SimulationSRE5_20.png}
%		\caption{Outliers in $5\%p$ and $20\%n$.}
%		\label{fig:simSRE5.20} 
%	\end{subfigure}
%	\vspace{0.4cm}
	\begin{subfigure}{.5\textwidth}
		\centering
		\includegraphics[width=.99\linewidth]{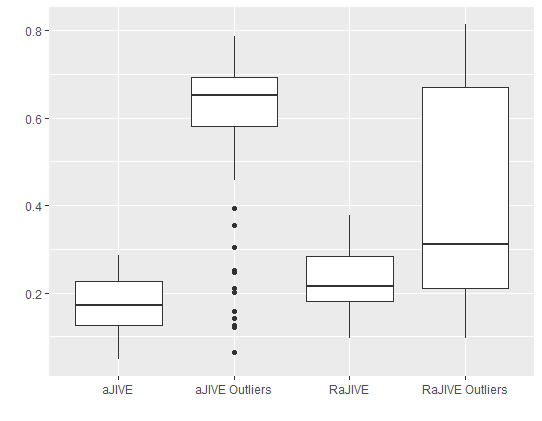}
		\caption{Outliers in $10\%$ variables, Set A.}
		\label{fig:simSRE10_SetA}
	\end{subfigure}
\vspace{0.4cm}
	\begin{subfigure}{.5\textwidth}
	\centering
	\includegraphics[width=.99\linewidth]{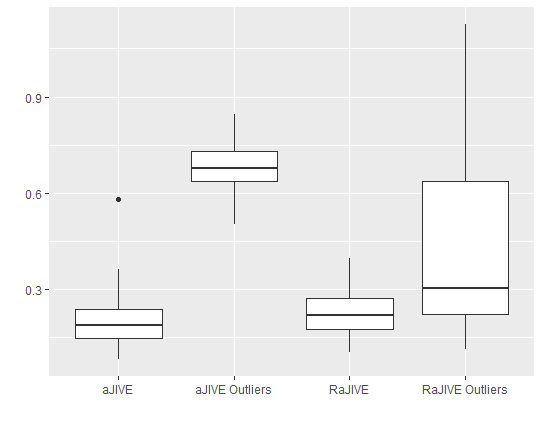}
	\caption{Outliers in $5\%$ variables, Set B.}
	\label{fig:simSRE5.10_SetB}
\end{subfigure}%
\begin{subfigure}{.5\textwidth}
	\centering
	\includegraphics[width=.99\linewidth]{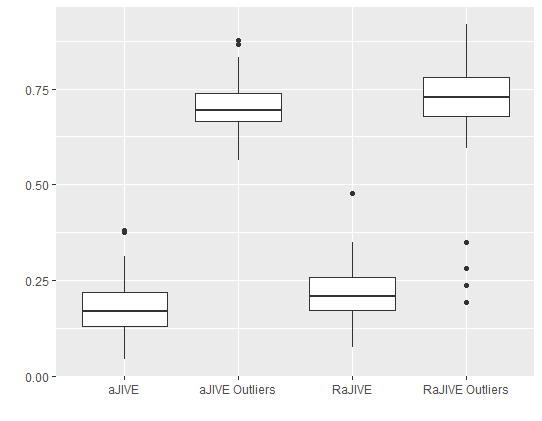}
	\caption{Outliers in $10\%$ variables, Set B.}
	\label{fig:simSRE10_SetB}
\end{subfigure}%
	\caption{Subspace recovery error of aJIVE and RaJIVE in simulation studies (Set A and Set B) 
		with outliers in different proportion of variables  and $10\%$ samples.} 	
	\label{fig:simSRE}
\end{figure}

\noindent\textbf{Results:} 
Table \ref{tab:simRanks} and Figures \ref{fig:simVar_SetA}--\ref{fig:simSRE} illustrate the performance measures 
obtained by both aJIVE and RaJIVE in the simulation studies (Set A--B) with and without outliers. 
Note that, when outliers are generated in $5\%$ of the variables as per either Set A or B, 
RaJIVE performs well and retrieved the correct joint ranks (Table \ref{tab:simRanks}a)), 
as well as a correct estimation of the proportion of variance explained in most cases (Figures \ref{fig:simVar5.10_SetA}, \ref{fig:simVar5.10_SetB}). 
While the standard aJIVE has a very high subspace recovery error in the presence of outliers, 
that of RaJIVE stays significantly lower (Figures \ref{fig:simSRE5.10_SetA}, \ref{fig:simSRE5.10_SetB}).
When affecting $10\%$ of variables with outliers as per Set A, the RaJIVE estimates still remains robust
and better than the usual aJIVE under data contamination (Table \ref{tab:simRanks}b); Figures \ref{fig:simVar10_SetA}, \ref{fig:simSRE10_SetB}). 
However, when outliers are added to $10\%$ of variables in Set B, 
the quality of the RaJIVE estimates deteriorates (Table \ref{tab:simRanks}b; Figures \ref{fig:simVar10_SetA}, \ref{fig:simSRE10_SetB}),
due to the presence of significantly higher amount of contamination (more than what it can tolerate).

\begin{table}[h!]
	\centering
	\begin{tabular}{c| c c c c}
		a) & & Outliers in $5\%$ of p & & \\
		Method & Joint & Ind Source 1 & Ind Source 2 & Ind Source 3 \\ \hline
\multicolumn{5}{c}{\underline{Set A}}\\
aJIVE & 3 & 17 & 9 & 5 \\
RaJIVE & 3 & 17  & 10 & 5 \\
aJIVE with outliers & 4 & 17 & 9 & 4\\
RaJIVE with outliers & 3 & 18 &10 & 5\\ 
\multicolumn{5}{c}{\underline{Set B}}\\
		aJIVE & 3 & 8 & 9 & 5 \\
		RaJIVE & 3 & 8 & 9 & 5 \\
		aJIVE with outliers & 4 & 7 & 9 & 4\\
		RaJIVE with outliers & 3 & 7 & 9 & 5\\ \hline
\multicolumn{5}{c}{} \\
		%	b) & & Outliers in $5\%p$ and $20\%n$ & & \\
		%	Method & Joint & Ind Source 1 & Ind Source 2 & Ind Source 3 \\ \hline
		%	aJIVE & 3 & 8 & 9 & 5 \\
		%	RaJIVE & 3 & 8  & 9 & 5 \\
		%	aJIVE with outliers & 4 & 7 & 9 & 4\\
		%	RaJIVE with outliers & 4 & 7 & 9 & 4\\
		%	\vspace{0.4cm} \\
		b) & & Outliers in $10\%$ of p  & & \\
		Method & Joint & Ind Source 1 & Ind Source 2 & Ind Source 3 \\ \hline
\multicolumn{5}{c}{\underline{Set A}}		\\
		aJIVE & 3 & 17 & 9 & 5 \\
RaJIVE & 3 & 17  & 10 & 5 \\
aJIVE with outliers & 4 & 17 & 9 & 4\\
RaJIVE with outliers & 3 & 17 & 9 & 5\\
\multicolumn{5}{c}{\underline{Set B}}\\
aJIVE & 3 & 7 & 9 & 5 \\
		RaJIVE & 3 & 8  & 9 & 5 \\
		aJIVE with outliers & 4 & 7 & 9 & 4\\
		RaJIVE with outliers & 4 & 7 & 9 & 4\\\hline
		%		\vspace{0.4cm} \\
		%	d) & & Outliers in $10\%p$ and $20\%n$ & & \\
		%	Method & Joint & Ind Source 1 & Ind Source 2 & Ind Source 3 \\ \hline
		%	aJIVE & 3 & 7 & 9 & 5 \\
		%	RaJIVE & 3 & 8  & 10 & 5 \\
		%	aJIVE with outliers & 4 & 7 & 9 & 4\\
		%	RaJIVE with outliers & 4 & 7 & 9 & 4\\
	\end{tabular}
	\caption{Median joint and individual ranks estimated by aJIVE and RaJIVE in simulation studies with outliers in different proportions of variables.}
	\label{tab:simRanks}	
\end{table}

\newpage
\section{Real-life application: TCGA breast cancer data}

Finally we illustrate applicability of our proposed RaJIVE to a real-life dataset, from TCGA, on breast cancer samples. 
The data preprocessing is described in \cite{lock.dunson2013} and used in \cite{oconnell.etal2016}. 
It contains the records for 348 breast tumor samples relative to three sources of omics data, 
namely gene expression for 654 genes, DNA methylation for 574 CpG sites and miRNA expression for 423 miRNAs. 
To examine the robustness of our procedure, we further contaminate $5\%$ of the variables and $10\%$ of the data points, 
by adding normally distributed outliers with mean $\mu_c = 3m+5s$ and standard deviation $\sigma_c = 3s$, 
where $m$ and $s$ are the mean and standard deviation of the affected variable (as in Set B of the simulation study).  
We compared the joint and individual ranks obtained by the standard aJIVE and the robust aJIVE, 
as well as the proportions of explained variation estimated by the two methods, both with and without data contamination.

\begin{figure}
	\centering
	\includegraphics[width=.99\linewidth]{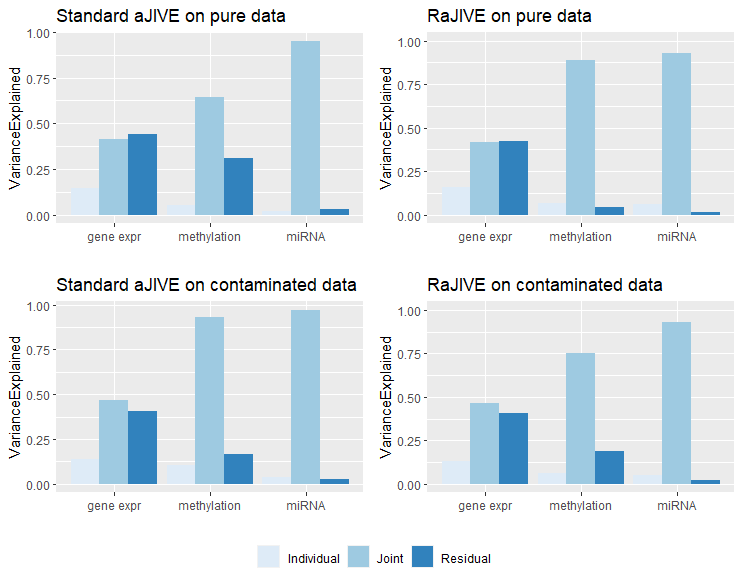}
	\caption{Proportion of variance explained by the aJIVE and robust aJIVE (RaJIVE) components 
		in the TCGA breast cancer data with and without data contamination.}	\label{fig:var}
\end{figure}

\begin{figure}
	\centering
	\begin{subfigure}{.5\textwidth}
		\centering
		\includegraphics[width=.95\linewidth]{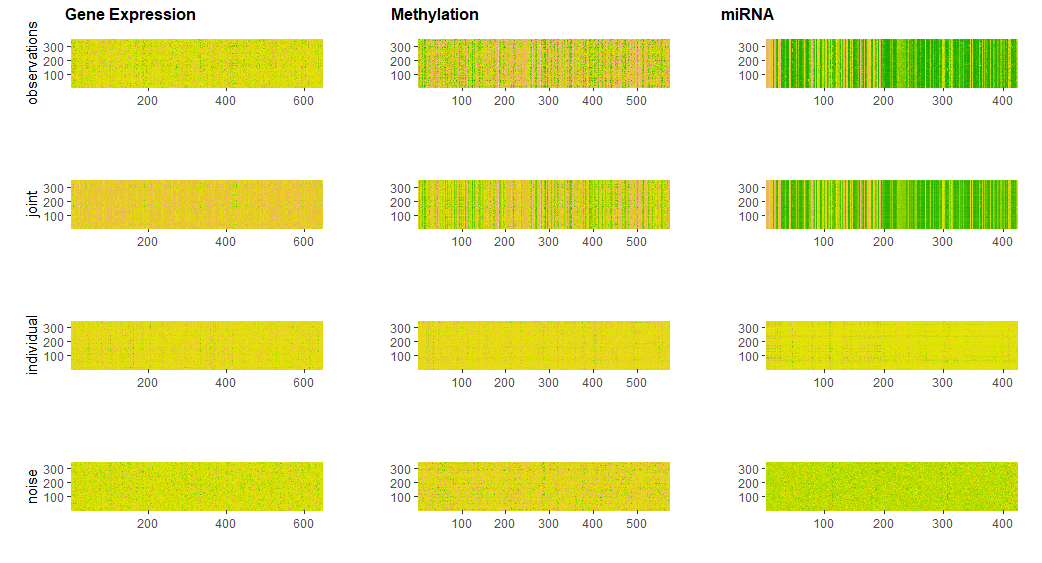}
		\caption{aJIVE decomposition on pure data}
		\label{fig:hmAS}
	\end{subfigure}%
	\begin{subfigure}{.5\textwidth}
		\centering
		\includegraphics[width=.95\linewidth]{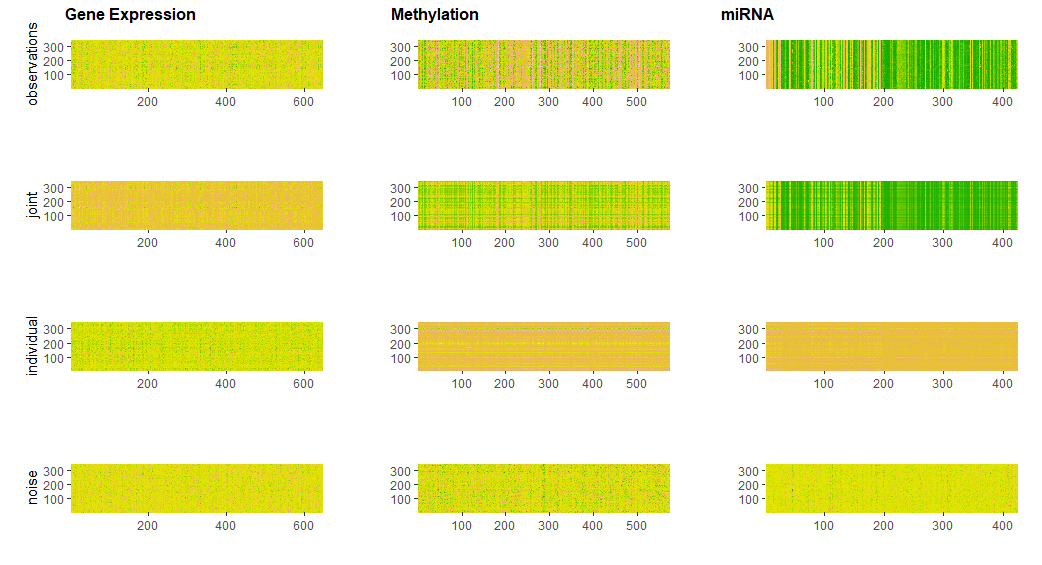}
		\caption{RaJIVE decomposition on pure data}
		\label{fig:hmRS}
	\end{subfigure}
	
	\begin{subfigure}{.5\textwidth}
		\centering
		\includegraphics[width=.95\linewidth]{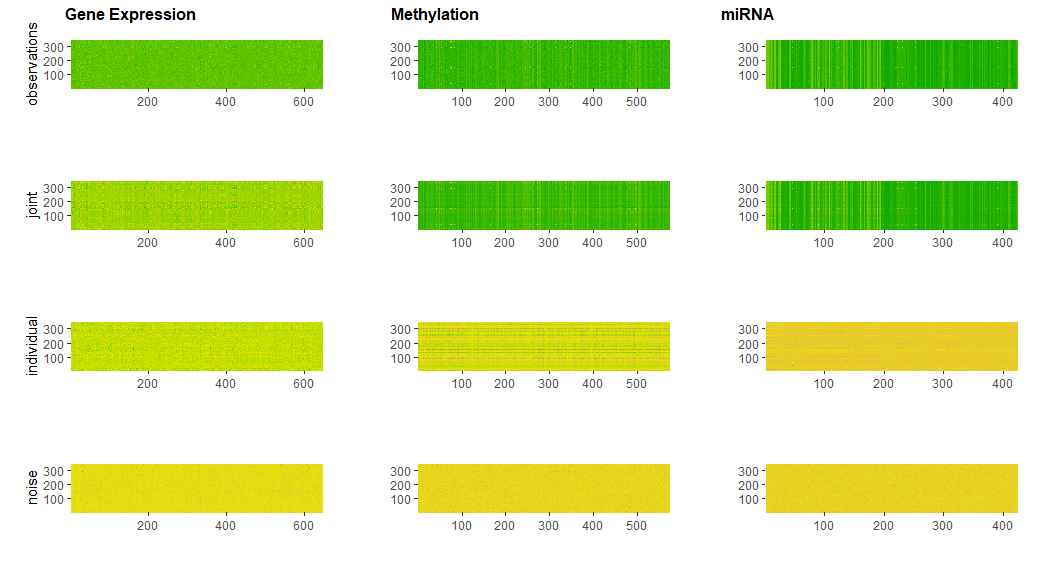}
		\caption{aJIVE decomposition on contaminated data}
		\label{fig:hmAO}
	\end{subfigure}%
	\begin{subfigure}{.5\textwidth}
		\centering
		\includegraphics[width=.95\linewidth]{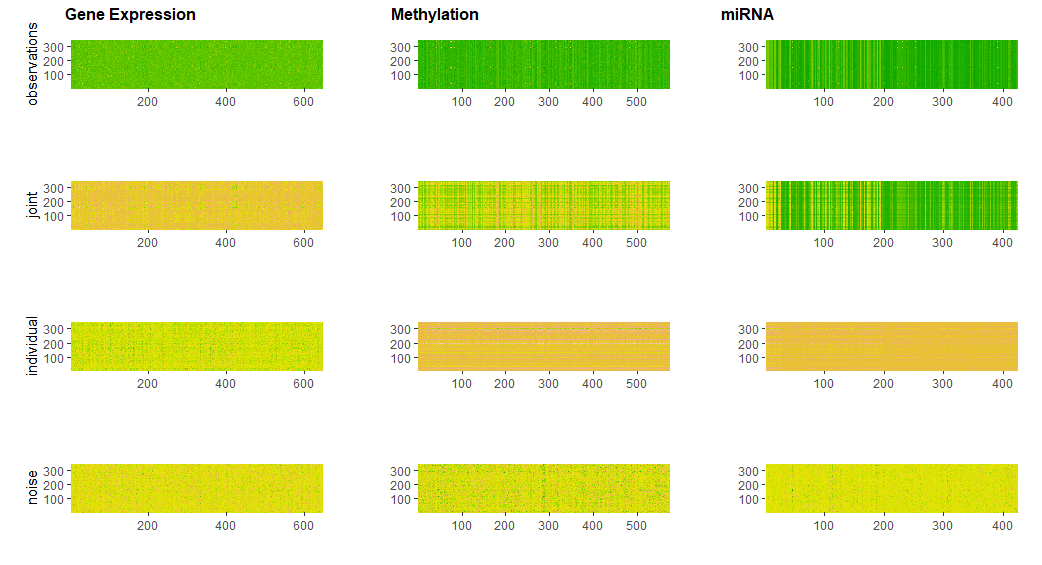}
		\caption{RaJIVE decomposition on contaminated data}
		\label{fig:hmRO}
	\end{subfigure}

	\caption{Heatmaps of the standard aJIVE and proposed RaJIVE decompositions in the TCGA breast cancer data.}	\label{fig:hm}
\end{figure}

Joint and individual ranks estimated by aJIVE and RaJIVE in the pure (with no outliers) and contaminated (with outliers) data are reported in Table \ref{tab:TCGAranks}. Although robust aJIVE is not able to recover the correct ranks, it is successful in estimating the correct proportion of variance explained (Figure \ref{fig:var}). Also the heatmap decompositions show that, while the aJIVE decomposition is extremely sensitive to outliers, RaJIVE is able to estimate the correct decomposition in presence of outliers (Figure \ref{fig:hm}).
\begin{table}[h!]
	\centering
	\begin{tabular}{c| c c c c c}
		Method & Data & Joint & Ind Expression & Ind Methylation & Ind miRNA \\ \hline
		Standard aJIVE & pure  & 11 & 12 & 5 & 10 \\
		Robust aJIVE & pure & 10 &  14 & 6  & 10 \\
		Standard aJIVE & contaminated & 10 & 12 & 6 & 11 \\
		Robust aJIVE & contaminated & 10 & 13  & 7  &  11\\
		
	\end{tabular}
	\caption{Joint and individual ranks obtained by aJIVE and RaJIVE.}
	\label{tab:TCGAranks}
	
\end{table}

\section{Discussion}

We propose a robust alternative to the aJIVE method for the estimation of joint and individual components 
in the presence of outliers in multi-source data. 
Our method (RaJIVE) uses a robust formulation of the SVD within the standard aJIVE algorithm. 
The performance of this proposed RaJIVE is illustrated in a set of simulation studies 
and on real data from The Cancer Genome Atlas, specifically on multi-platform omics data collected on breast cancer samples. 

We investigate, in detail, the effect of increasing proportions of outliers on the joint and individual components identified by aJIVE. 
Interestingly, our simulation study shows that the presence of outliers has an effect on the joint and individual components 
only when all the data sources are affected, and that the effect is negligible when only one source shows the presence of outliers, 
regardless of the proportion of affected variables. 
This appears very relevant when focusing our interest on the joint contributions of the data layers. 

A major issue related to the aJIVE method is the selection of initial ranks, most commonly based on the visualization of screeplots, 
which is highly subjective and sensitive to noises in the data. 
Although an alternative based on the profile likelihood idea was suggested \cite{zhu.ghodsi2006}, 
this addresses the problem only partially and still remains quite subjective. 
This issue is likely to affect the proposed RaJIVE as well, and the initial ranks need to be selected carefully before implementing both methods. 
We believe this to be a possible reason for the incorrect estimation of the ranks in the TCGA application. 

Although we only consider normally distributed noise to generate outliers, 
the robust SVD formulation used in RaJIVE accounts for different types of noise in the data \citep{zhang.etal2013}. 
Therefore, we expect the proposed RaJIVE to be useful in situations with other types of error as well, and also in missing data situations.
However, it has been observed through simulations that RaJIVE has a noise-tolerance limit (breakdown-point) 
and it may gets affected by the presence of contamination over this limit.  
It would be extremely important in future work to overcome this limitation of the proposed RaJIVE,
possibly by using a more robust SVD computation procedure that would be able to produce stable results even under heavy contaminations.

\subsection*{Data and software availability}
A \texttt{R} package with the implementation of RaJIVE is available at \href{https://github.com/ericaponzi/RaJIVE}{https://github.com/ericaponzi/RaJIVE}.

\subsection*{Funding}
Norwegian Research Council - grant number 248804: National training initiative to make better use of biobanks and health registry data.
Research of AG is also supported by an INSPIRE Faculty research grant and a grant (No.~SRG/2020/000072) from 
Science and Engineering Research Board, both under the Department of Science and Technology, Government of India. 

%\subsection*{Acknowledgments}

\appendix
\section{Computation of the robRSVD}
\label{App1}

The computation of the robRSVD is based on the alternating regression formulation of the SVD problem. 
It estimates one singular value and the corresponding singular vectors at a time.
The original implementation of robRSVD in the R-package `\texttt{RobRSVD}' \cite{zhang.etal2013}
additionally considers the regularization of the singular vectors, which is not necessary in our context.
An appropriately modified algorithm for computation of the robRSVD is used in our proposed RaJIVE to increase computational efficiency,
which is described here.

Suppose that we wish to find the SVD of a matrix $\bm{X}$ of order $m\times n$. 
In robRSVD, we first compute the rank-one approximation of the matrix as 
$\bm{X}\approx\bm{a}\bm{b}^T$ for some vectors $\bm{a}=(a_1, \ldots, a_m)^T$ and $\bm{b}=(b_1, \ldots, b_n)^T$. 
%Now, to compute the estimates of 
To estimate $\bm{a}$ and $\bm{b}$, we re-express this problem of matrix decomposition as the linear regression problem
given by 
\begin{eqnarray}
x_{ij} = a_ib_j + e_{ij}, ~~~~~~i=1, \ldots, m; ~j=1, \ldots, n,
\label{EQ:svd_reg}
\end{eqnarray}
where $x_{ij}$ denotes the $(i, j)$-th element of matrix $\bm{X}$ and $e_{ij}$ are residual (random error) components for all $i, j$. 
If we fix the values of $a_i$ and any one index $j$, we get a linear regression problem in (\ref{EQ:svd_reg}) with coefficient $b_j$
and $m$ observations (by varying $i$) and any standard estimation method can be used to estimate the corresponding regression coefficients $b_j$. 
Varying over all $j$, we get estimates of $\boldsymbol{b}=(b_1, \ldots, b_n)$ given $\bm{a}$. 
Next, given these estimated values of $b_j$s, we treat them as covariates in (\ref{EQ:svd_reg}) and estimate the $a_i$s as the unknown parameter
in $m$ linear regressions for varying $i=1, \ldots, m$ (with $n$ observations, varying $j$, for each $i$). 
These two steps of regressions approach can be repeated alternatively, 
with the new estimated parameter values being used in each steps, until convergence to get the final estimates of $\bm{a}$ and $\bm{b}$.

In the usual SVD, in each step of the alternating regression as described above, 
the parameter estimation is done by the standard least-squares approach. 
In robRSVD, however, the regression coefficients are estimated by the robust M-estimation approach \citep{huber.ronchetti2011}
which is computed via an appropriate reweighted least-squares algorithm. 
More specifically, 
%Huber's weight function is used to downweight the effects of large residuals associated with the outlier (noisy) observations.
it converts the estimation of the parameters $\bm{a}$ and $\bm{b}$ in the above alternative regression approach
to an optimization problem as
\begin{eqnarray}
(\widehat{\bm{a}}, \widehat{\bm{b}}) = \arg\min\limits_{(\bm{a},\bm{b})} \rho\left(\frac{\bm{X}-\bm{a}\bm{b}^T}{\sigma}\right),
\label{EQ:RobRSVD}
\end{eqnarray}
where $\sigma^2$ is the (assumed) common variance of $e_{ij}$s and $\rho$ is the Huber's loss function given by 
$$
\rho(x) = \left\{ \begin{array}{ll}
x^2    & \mbox{ if }|x|\leq c,
\\
2c|x|-c^2    & \mbox{ if }|x|> c.
\end{array}\right.
$$
Here $c>0$ is a robustness tuning parameter that controls the balance between efficiency and robustness of the resulting estimators;
the suggested choice of $c$ is $1.345$ which provides 95\% efficiency under a linear regression model with normal errors \citep{huber.ronchetti2011}. 
Solving the optimization problem (\ref{EQ:RobRSVD}) numerically via iterative reweighted  least-squares algorithm,
we get the robust estimates $(\widehat{\bm{a}}, \widehat{\bm{b}})$ of $(\bm{a}, \bm{b})$ in RobRSVD.
Then, a singular value of $\bm{X}$ is  obtained as
$\delta = ||\widehat{\bm{a}}|| ||\widehat{\bm{b}}||$, with $||\cdot||$ denoting the Euclidean norm, 
and the corresponding normalized left and right singular vectors as $\bm{u}=\widehat{\bm{a}}/||\widehat{\bm{a}}||$ 
and $\bm{v}=\widehat{\bm{b}}/||\widehat{\bm{b}}||$, respectively.

%and the corresponding singular vectors $\bm{u}, \bm{v}$ in the robRSVD, as described previously. 
%Once their robust estimates, say $\widehat{\bm{a}}$ 
%and $\widehat{\bm{b}}$ are obtained, we get one singular value of $\bm{X}$ as 

Once the first singular value and vectors for $\bm{X}$ are obtained, in robRSVD, 
the second one needs to be obtained by applying the same procedure on the residual matrix $\bm{X}_1 = \bm{X} - \delta \bm{u}\bm{v}^T$.
Repeating this procedure sequentially, we get all the singular values of $\bm{X}$ and the corresponding (normalized) singular vectors.

%\bibliographystyle{plain}

%\section*{Figures}

%%%%%%%%%%%%%%%%%%%%%%%%%%%%%%%%%%%%%%%%%%%%%%%%% RaJIVE & aJIVE %%%%%%%%%%%%%%%%%%%%%%%%%%%%%

%%%%%%%%%%%%%%%%%%%%%%%%%%%%%%%%%%%%%%%%%%%%%%%%% Data Analysis %%%%%%%%%%%%%%%%%%%%%%%%%%%%%

\bibliography{MainText}

\begin{thebibliography}{}

\bibitem[Carmichael, 2019]{ajiveR}
Carmichael, I. (2019).
\newblock ajive: Angle based joint and individual variation explained.
\newblock \url{https://github.com/idc9/r_jive}.

\bibitem[Fan et~al., 2019]{fan.etal2019}
Fan, J., Wang, D., Wang, K., and Zhu, Z. (2019).
\newblock Distributed estimation of principal eigenspaces.
\newblock {\em Annals of Statistics}, 47:3009--3031.

\bibitem[Feng et~al., 2018]{FENG2018241}
Feng, Q., Jiang, M., Hannig, J., and Marron, J. (2018).
\newblock Angle-based joint and individual variation explained.
\newblock {\em Journal of Multivariate Analysis}, 166:241 -- 265.

\bibitem[Hawkins et~al., 2001]{Hawkins.etal2001}
Hawkins, D., Liu, L., and Young, S.~S. (2001).
\newblock Robust singular value decomposition, technical report number 122.
\newblock In {\em National Institute of Statistical Sciences 19}.

\bibitem[Hellton and Thoresen, 2016]{hellton.thoresen2016}
Hellton, K.~H. and Thoresen, M. (2016).
\newblock {Integrative clustering of high-dimensional data with joint and
  individual clusters}.
\newblock {\em Biostatistics}, 17(3):537--548.

\bibitem[Huang et~al., 2017]{huang.etal2017}
Huang, S., Chaudhary, K., and Garmire, L.~X. (2017).
\newblock More is better: Recent progress in multi-omics data integration
  methods.
\newblock {\em Frontiers in Genetics}, 8.

\bibitem[Huber and Ronchetti, 2011]{huber.ronchetti2011}
Huber, P. and Ronchetti, E. (2011).
\newblock {\em Robust Statistics}.
\newblock Wiley Series in Probability and Statistics. Wiley.

\bibitem[Jiang, 2018]{ajiveM}
Jiang, M. (2018).
\newblock {AJIVE} project.
\newblock \url{https://github.com/MeileiJiang/AJIVE_Project}.

\bibitem[Kaplan and Lock, 2017]{kaplan.lock2017}
Kaplan, A. and Lock, E.~F. (2017).
\newblock Prediction with dimension reduction of multiple molecular data
  sources for patient survival.
\newblock {\em Cancer Inform}, 16:1--11.

\bibitem[Kuligowski et~al., 2015]{kuligowski.etal2015}
Kuligowski, J., Perez-Guaita, D., Sanchez-Illana, A. ad Leon-Gonzalez, Z.,
  de~la Guardia, M., Vento, M., Lock, E.~F., and Quintas, G. (2015).
\newblock Analysis of multi-source metabolomic data using joint and individual
  variation explained ({JIVE}).
\newblock {\em Analyst}, 13:4521--4529.

\bibitem[Lock and Dunson, 2013]{lock.dunson2013}
Lock, E.~F. and Dunson, D.~B. (2013).
\newblock Bayesian consensus clustering.
\newblock {\em Bioinformatics}, 29:2610--2616.

\bibitem[Lock et~al., 2013]{lock.etal2013}
Lock, E.~F., Hoadley, K.~A., Marron, J.~S., and Nobel, A.~B. (2013).
\newblock Joint and individual variation explained ({JIVE}) for integrated
  analysis of multiple data types.
\newblock {\em Annals of Applied Statistics}, 7:523--542.

\bibitem[Lofsted et~al., 2012]{lofsted.etal2012}
Lofsted, T., Hoffman, D., and Trygg, J. (2012).
\newblock Global, local and unique decomposition in {O}n{PLS} for multiblock
  data analysis.
\newblock {\em Analytica Chimica Acta}, 791:13--24.

\bibitem[M{\aa}ge et~al., 2019]{mage.etal2019}
M{\aa}ge, I., Smilde, A.~K., and van~der Kloet, F.~M. (2019).
\newblock Performance of methods that separate common and distinct variation in
  multiple data blocks.
\newblock {\em Journal of Chemometrics}, 33:e3085.

\bibitem[O`Connell and Lock, 2016]{oconnell.etal2016}
O`Connell, M.~J. and Lock, E.~F. (2016).
\newblock {R.JIVE} for exploration of multi-source molecular data.
\newblock {\em Bioinformatics}, 32(18):2877--2879.

\bibitem[Panagakis et~al., 2016]{panagakis.etal2016}
Panagakis, Y., Nicolau, M.~A., Zafeiriou, S., and Pantic, M. (2016).
\newblock Robust correlated and individual component analysis.
\newblock {\em IEEE Transactions on Pattern Analysis and Machine Intelligence},
  38:1665--1678.

\bibitem[Ponzi et~al., 2020]{ponzi.etal2020}
Ponzi, E., Thoresen, M., N{\o}st, T.~H., and M{\o}llersen, K. (2020).
\newblock Integrative analysis of multi-omics data improves model predictions:
  an application to lung cancer.
\newblock {\em bioRxiv}, 10.

\bibitem[Rappaport and Ron, 2018]{rappaport.ron2018}
Rappaport, N. and Ron, S. (2018).
\newblock Multi-omic and multi-view clustering algorithms: review and cancer
  benchmark.
\newblock {\em Nucleic acids research}, 42:10546--10562.

\bibitem[Sagonas et~al., 2017]{sagonas.etal2017}
Sagonas, C., Panagakis, Y., Leidinger, A., and Zafeiriou, S. (2017).
\newblock Robust joint and individual variance explained.
\newblock In {\em 2017 IEEE Conference on Computer Vision and Pattern
  Recognition (CVPR)}.

\bibitem[Schouteden et~al., 2013]{schouteden.etal2013}
Schouteden, M., Van~Deun, T.~F., Wilderjans, T., and Van~Mechelen, I. (2013).
\newblock Performing {DISCO-SCA} to search for distinctive and common
  information in linked data.
\newblock {\em Behavior Research Methods}, 46:576--587.

\bibitem[Tang and Allen, 2018]{tang2018integrated}
Tang, T.~M. and Allen, G.~I. (2018).
\newblock Integrated principal components analysis.

\bibitem[Tseng et~al., 2015]{tseng.etal2015}
Tseng, G., Ghosh, D., and Zhou, X.~J. (2015).
\newblock {\em Integrating omics data}.
\newblock Cambridge University Press.

\bibitem[Zhang and Pan, 2013]{RobRpackage}
Zhang, L. and Pan, C. (2013).
\newblock {\em RobRSVD: Robust Regularized Singular Value Decomposition}.

\bibitem[Zhang et~al., 2013]{zhang.etal2013}
Zhang, L., Shen, H., and Huang, J.~Z. (2013).
\newblock Robust regularized singular value decomposition with application to
  mortality data.
\newblock {\em The Annals of Applied Statistics}, 7:1540--1561.

\bibitem[Zhu and Ghodsi, 2006]{zhu.ghodsi2006}
Zhu, M. and Ghodsi, A. (2006).
\newblock Automatic dimensionality selection from the scree plot via the use of
  profile likelihood.
\newblock {\em Computational Statistics and Data Analysis}, 51:918--930.

\end{thebibliography}
\end{document}